\documentstyle[twoside,psfig]{article}

\def\be{\begin{equation}}
\def\ee{\end{equation}}
\def\bea{\begin{eqnarray}}
\def\eea{\end{eqnarray}}

\catcode`\@=11
\long\def\@makefntext#1{
\protect\noindent \hbox to 3.2pt {\hskip-.9pt
$^{{\eightrm\@thefnmark}}$\hfil}#1\hfill}               %CAN BE USED

\def\@makefnmark{\hbox to 0pt{$^{\@thefnmark}$\hss}}    %ORIGINAL

\def\ps@myheadings{\let\@mkboth\@gobbletwo
\def\@oddhead{\hbox{}
\rightmark\hfil\eightrm\thepage}
\def\@oddfoot{}\def\@evenhead{\eightrm\thepage\hfil
\leftmark\hbox{}}\def\@evenfoot{}
\def\sectionmark##1{}\def\subsectionmark##1{}}

\oddsidemargin=\evensidemargin
\newcounter{sectionc}\newcounter{subsectionc}\newcounter{subsubsectionc}
\renewcommand{\section}[1] {\vspace{12pt}\addtocounter{sectionc}{1}
\setcounter{subsectionc}{0}\setcounter{subsubsectionc}{0}\noindent
        {\tenbf\thesectionc. #1}\par\vspace{5pt}}
\renewcommand{\subsection}[1] {\vspace{12pt}\addtocounter{subsectionc}{1}
        \setcounter{subsubsectionc}{0}\noindent
        {\bf\thesectionc.\thesubsectionc. {\kern1pt \bfit #1}}\par\vspace{5pt}}
\renewcommand{\subsubsection}[1] {\vspace{12pt}\addtocounter{subsubsectionc}{1}
        \noindent{\tenrm\thesectionc.\thesubsectionc.\thesubsubsectionc.
        {\kern1pt \tenit #1}}\par\vspace{5pt}}
\newcommand{\nonumsection}[1] {\vspace{12pt}\noindent{\tenbf #1}
        \par\vspace{5pt}}

\newcounter{appendixc}
\newcounter{subappendixc}[appendixc]
\newcounter{subsubappendixc}[subappendixc]
\renewcommand{\thesubappendixc}{\Alph{appendixc}.\arabic{subappendixc}}
\renewcommand{\thesubsubappendixc}
        {\Alph{appendixc}.\arabic{subappendixc}.\arabic{subsubappendixc}}

\renewcommand{\appendix}[1] {\vspace{12pt}
        \refstepcounter{appendixc}
        \setcounter{figure}{0}
        \setcounter{table}{0}
        \setcounter{lemma}{0}
        \setcounter{theorem}{0}
        \setcounter{corollary}{0}
        \setcounter{definition}{0}
        \setcounter{equation}{0}
        \renewcommand{\thefigure}{\Alph{appendixc}.\arabic{figure}}
        \renewcommand{\thetable}{\Alph{appendixc}.\arabic{table}}
        \renewcommand{\theappendixc}{\Alph{appendixc}}
        \renewcommand{\thelemma}{\Alph{appendixc}.\arabic{lemma}}
        \renewcommand{\thetheorem}{\Alph{appendixc}.\arabic{theorem}}
        \renewcommand{\thedefinition}{\Alph{appendixc}.\arabic{definition}}
        \renewcommand{\thecorollary}{\Alph{appendixc}.\arabic{corollary}}
        \renewcommand{\theequation}{\Alph{appendixc}.\arabic{equation}}
        \noindent{\tenbf Appendix \theappendixc #1}\par\vspace{5pt}}
\newcommand{\subappendix}[1] {\vspace{12pt}
        \refstepcounter{subappendixc}
        \noindent{\bf Appendix \thesubappendixc. {\kern1pt \bfit #1}}
        \par\vspace{5pt}}
\newcommand{\subsubappendix}[1] {\vspace{12pt}
        \refstepcounter{subsubappendixc}
        \noindent{\rm Appendix \thesubsubappendixc. {\kern1pt \tenit #1}}
        \par\vspace{5pt}}

\newcommand{\textlineskip}{\baselineskip=13pt}
\newcommand{\smalllineskip}{\baselineskip=10pt}

\def\eightcirc{
\begin{picture}(0,0)
\put(4.4,1.8){\circle{6.5}}
\end{picture}}
\def\eightcopyright{\eightcirc\kern2.7pt\hbox{\eightrm c}}

\newcommand{\copyrightheading}[1]
        {\vspace*{-2.5cm}\smalllineskip{\flushleft
        {\footnotesize Modern Physics Letters A #1}\\
        {\footnotesize $\eightcopyright$\, World Scientific Publishing
         Company}\\
         }}

\def\abstracts#1#2#3{{
        \centering{\begin{minipage}{4.5in}\footnotesize\baselineskip=10pt
        \parindent=0pt #1\par
        \parindent=15pt #2\par
        \parindent=15pt #3
        \end{minipage}}\par}}

\def\keywords#1{{
        \centering{\begin{minipage}{4.5in}\footnotesize\baselineskip=10pt
        {\footnotesize\it Keywords}\/: #1
         \end{minipage}}\par}}

\def\pacs#1{{
        \centering{\begin{minipage}{4.5in}\baselineskip=10pt
        {PACS NO(s)}\/: #1
         \end{minipage}}\par}}

\newcommand{\bibit}{\nineit}
\newcommand{\bibbf}{\ninebf}
\renewenvironment{thebibliography}[1]
        {\frenchspacing
         \ninerm\baselineskip=11pt
         \begin{list}{\arabic{enumi}.}
        {\usecounter{enumi}\setlength{\parsep}{0pt}
         \setlength{\leftmargin 12.7pt}{\rightmargin 0pt} %FOR 1--9 ITEMS
         \setlength{\itemsep}{0pt} \settowidth
        {\labelwidth}{#1.}\sloppy}}{\end{list}}

\newcounter{itemlistc}
\newcounter{romanlistc}
\newcounter{alphlistc}
\newcounter{arabiclistc}

\newenvironment{romanlist}
        {\setcounter{romanlistc}{0}
         \begin{list}{$($\roman{romanlistc}$)$}
        {\usecounter{romanlistc}
         \setlength{\parsep}{0pt}
         \setlength{\itemsep}{0pt}}}{\end{list}}

\newcommand{\fcaption}[1]{
        \refstepcounter{figure}
        \setbox\@tempboxa = \hbox{\footnotesize Fig.~\thefigure. #1}
        \ifdim \wd\@tempboxa > 5in
           {\begin{center}
        \parbox{5in}{\footnotesize\smalllineskip Fig.~\thefigure. #1}
            \end{center}}
        \else
             {\begin{center}
             {\footnotesize Fig.~\thefigure. #1}
              \end{center}}
        \fi}

\newcommand{\tcaption}[1]{
        \refstepcounter{table}
        \setbox\@tempboxa = \hbox{\footnotesize Table~\thetable. #1}
        \ifdim \wd\@tempboxa > 5in
           {\begin{center}
        \parbox{5in}{\footnotesize\smalllineskip Table~\thetable. #1}
            \end{center}}
        \else
             {\begin{center}
             {\footnotesize Table~\thetable. #1}
              \end{center}}
        \fi}

\def\@citex[#1]#2{\if@filesw\immediate\write\@auxout
        {\string\citation{#2}}\fi
\def\@citea{}\@cite{\@for\@citeb:=#2\do
        {\@citea\def\@citea{,}\@ifundefined
        {b@\@citeb}{{\bf ?}\@warning
        {Citation `\@citeb' on page \thepage \space undefined}}
        {\csname b@\@citeb\endcsname}}}{#1}}

\newif\if@cghi
\def\cite{\@cghitrue\@ifnextchar [{\@tempswatrue
        \@citex}{\@tempswafalse\@citex[]}}
\def\citelow{\@cghifalse\@ifnextchar [{\@tempswatrue
        \@citex}{\@tempswafalse\@citex[]}}
\def\@cite#1#2{{$\null^{#1}$\if@tempswa\typeout
        {IJCGA warning: optional citation argument
        ignored: `#2'} \fi}}

\def\pmb#1{\setbox0=\hbox{#1}
        \kern-.025em\copy0\kern-\wd0
        \kern.05em\copy0\kern-\wd0
        \kern-.025em\raise.0433em\box0}

\def\fnt#1#2{\footnotetext{\kern-.3em
        {$^{\mbox{\scriptsize #1}}$}{#2}}}

\def\ps@myheadings{%
    \let\@oddfoot\@empty\let\@evenfoot\@empty
    \def\@evenhead{\slshape\leftmark\hfil}%       %EVEN PAGE
    \def\@oddhead{\hfil{\slshape\rightmark}}%     %ODD PAGE
    \let\@mkboth\@gobbletwo
    \let\sectionmark\@gobble
    \let\subsectionmark\@gobble
    }
\font\tenrm=cmr10
\font\tenit=cmti10
\font\tenbf=cmbx10
\font\bfit=cmbxti10 at 10pt
\font\ninerm=cmr9
\font\nineit=cmti9
\font\ninebf=cmbx9
\font\eightrm=cmr8

\def\qed{\hbox{${\vcenter{\vbox{                        %HOLLOW SQUARE
   \hrule height 0.4pt\hbox{\vrule width 0.4pt height 6pt
   \kern5pt\vrule width 0.4pt}\hrule height 0.4pt}}}$}}

\begin{document}
\setlength{\textheight}{7.7truein}  %for 2nd page onwards

\thispagestyle{empty}

%\markboth{\protect{\footnotesize\it Instructions for Typesetting
%Manuscripts}}{\protect{\footnotesize\it Instructions for
%Typesetting Manuscripts}}

\normalsize\textlineskip

\setcounter{page}{1}

\copyrightheading{}     %{Vol.~0, No.~0 (2002) 000--000}

\vspace*{0.88truein}

\centerline{\bf WHAT CAN SNO NEUTRAL CURRENT RATE TEACH US}
\baselineskip=13pt
\centerline{\bf ABOUT THE SOLAR NEUTRINO ANOMALY}
%\vspace*{0.37truein}
\vspace*{0.4truein}
\centerline{\footnotesize ABHIJIT BANDYOPADHYAY, SANDHYA CHOUBEY, 
SRUBABATI GOSWAMI}
\baselineskip=12pt
\centerline{\footnotesize\it Saha Institute of Nuclear Physics
}
\baselineskip=10pt
\centerline{\footnotesize\it Bidhannager, Kolkata 700 064,
India}
%\vspace*{10pt}
\vspace*{12pt}

\centerline{\footnotesize D.P.ROY}
\baselineskip=12pt
\centerline{\footnotesize\it Tata Institute of Fundamental Research}
\baselineskip=10pt
\centerline{\footnotesize\it Homi Bhabha Road, Mumbai 400 005, India}
%\vspace*{0.225truein}
\vspace*{0.228truein}

%\publisher{(received date)}{(revised date)}

%\vspace*{0.21truein}
\vspace*{0.23truein}
\abstracts{We investigate how the anticipated 
neutral current rate from $SNO$ will
sharpen our understanding of the solar neutrino anomaly.  Quantitative
analyses are performed with representative values of this rate in the
expected range of $0.8 - 1.2$.  This would provide a $5 - 10 \ \sigma$
signal for $\nu_e$ transition into a state containing an active
neutrino component.  Assuming this state to be purely active one can
estimate both the $^8B$ neutrino flux and the $\nu_e$ survival
probability to a much higher precision than currently possible.
Finally the measured value of the $NC$ rate will have profound
implications for the mass and mixing parameters of the solar neutrino
oscillation solution.}{}{}

\vspace*{10pt}
\vspace*{10pt}

\vspace*{2pt}

\textlineskip                  %) USE THIS MEASUREMENT WHEN THERE IS
\vspace*{12pt}                 %) NO SECTION HEADING

\baselineskip=13pt              %) ACTUAL LEADING

A large number of experiments have observed anomalously low solar
neutrino flux $^{1,2,3,4}$ compared to the standard solar model (SSM)
prediction .$^5$  They are the radiochemical experiments on $Ga$ $^1$
and $Cl$ $^2$ targets as well as the water Cerenkov experiments from
Super Kamiokande (SK) $^3$ and Sudbury Neutrino Observatory ($SNO$) $^4$.
SK observes the emitted electron from elastic scattering
\be
\nu + e \rightarrow \nu + e,
\label{one}
\ee
while $SNO$ observes it from the charged current process
\be
\nu_e + d \rightarrow p + p + e
\label{two}
\ee
using a heavy water target.  Both the experiments probe the high
energy tail of the solar neutrino spectrum, which is dominated by the
$^8B$ neutrino flux.  The SK elastic scattering process (1) is
sensitive to $\nu_e$ via CC and $NC$ interactions; but it also has a
limited sensitivity to $\nu_{\mu,\tau}$ via the $NC$ interaction.  On
the other hand the $SNO$ experiment is looking separately at the $NC$
rate, which has equal sensitivity to all the active neutrino flavors,
via
\be
\nu + d \rightarrow \nu + p + n,
\label{three}
\ee
followed by the neutron capture by $NaCl$.  The resulting excited
state decays via $\gamma$ emission, which constitutes the $NC$ signal.
The $SNO$ experiment is expected to report its first $NC$ rate shortly,
corresponding to a data sample of $\sim 10^3$ events -- i.e. similar
statistical accuracy as their 1st CC data .$^4$  The purpose of our
work is to critically analyse how far this data will enhance our
understanding of the solar neutrino anomaly and its solution.  

By far the most plausible solution of the solar neutrino anomaly is in
terms of neutrino oscillation, and in particular the oscillation of
$\nu_e$ into another active flavor $\nu_a$, which can be any
combination of $\nu_\mu$ and $\nu_\tau$.  Indeed we have made
considerable progress in understanding the solar neutrino anomaly and
its oscillation solution over the past few months by combining the CC
rate from $SNO$ with the SK elastic scattering data $^{6,7,8,9,10}$.  
Firstly they
disfavour $\nu_e$ transition into a sterile neutrino $\nu_s$ at the
$3\sigma$ level in a model independent way.$^4$  Secondly, assuming
only transition between $\nu_e$ and $\nu_a$, one can estimate both the
$^8B$ flux and the $\nu_e$ survival probability, although with fairly
large uncertainties.$^7$ Thirdly one can do a combined analysis of
the $SNO$, $SK$ and the earlier radiochemical data assuming SSM and the
$\nu_e \rightarrow \nu_a$ oscillation model .$^{7,8,9}$  One sees that the
result has narrowed down to only two possible solutions, both having
large mixing angles -- i.e. the so called LMA and LOW solutions.  We
shall see below that each of the above results can be significantly
sharpened further by including the $NC$ rate from $SNO$.

\begin{table}[h]  %[htbp]
\tcaption{The observed solar neutrino rates relative to the $SSM$
predictions are shown along with their compositions and threshold
energies for different experiments.  For the $SK$ experiment the
$\nu_e$ contribution to the rate $R$ is shown in parentheses assuming
$\nu_e \rightarrow \nu_a$ transition.}
\begin{tabular}{|c|c|c|c|}
\hline
&&& \\
experiment & $R$ & composition & E$_{th}$ \\
&&& (Mev)\\
\hline
&&& \\
$Ga$ & 0.584 $\pm$ 0.039 & $pp(55\%),Be(25\%),B(10\%)$ & 0.2 \\
&&& \\
$Cl$ & 0.335 $\pm$ 0.029 & $B(75\%),Be(15\%)$ & 0.8 \\
&&& \\
$SK$ & 0.459 $\pm$ 0.017 (0.351 $\pm$ 0.017) & $B(100\%)$ & 5.0 \\
&&& \\
$SNO(CC)$ & 0.347 $\pm$ 0.027 & $B(100\%)$ & 7.0 \\
&&& \\
\hline
\end{tabular}

\end{table}

Table 1 lists the observed solar neutrino rates of the $Ga,Cl,SK$ and
$SNO$ ($CC$) experiments relative to the corresponding $SSM$ predictions.
The compositions of the respective solar neutrino fluxes are also
indicated along with the threshold energies.  Assuming the SSM
neutrino fluxes and the transition of 
$\nu_e$ into an active flavor $\nu_a$ one can write the SK elastic
scattering rate relative to the SSM prediction in terms the survival
probability, i.e. 
\be
R^{el}_{SK} = P_{ee} + r P_{ea} = P_{ee} + r (1 - P_{ee}),
\label{four}
\ee
where $r = \sigma^{NC}_{\nu_a}/\sigma^{CC+NC}_{\nu_e} \simeq 0.17$ is
the ratio of $\nu_{\mu,\tau}$ to $\nu_e$ elastic scattering
cross-sections (\ref{one}).$^6$  The resulting value of the $\nu_e$
survival probability $P_{ee}$ is shown parenthetically in Table 1.
The other rates shown in this table are identical to the corresponding
$P_{ee}$, since they are sensitive to $\nu_e$ only.

As we see from table 1, the $SK$ and $SNO$ experiments are sensitive only
to the $^8B$ neutrino spectrum.  While the shape of this spectrum is
predicted with good precision by the SSM, there is a large
uncertainty in the predicted normalisation, $^5$ 
\be
\phi_B = 5.05 \times 10^6 \left(1^{+.20}_{-.16}\right) cm^{-2} s^{-1},
\label{five}
\ee
arising from the uncertainty in the $^7Be + p \rightarrow \ {^8B} + \gamma$
cross-section.  Therefore we shall introduce a constant parameter
$f_B$ to denote the normalisation of the $^8B$ neutrino flux relative
to the SSM prediction.  Then the SK elastic scattering and the $SNO$ $CC$
and $NC$ scattering rates relative to the corresponding $SSM$ predictions
are 
\bea
R^{el}_{SK} &=& f_B P_{ee} + f_B r P_{ea}, 
\label{six} \\[2mm]
R^{CC}_{SNO} &=& f_B P_{ee},
\label{seven} \\[2mm]
R^{NC}_{SNO} &=& f_B (P_{ee} + P_{ea}),
\label{eight}
\eea
which hold for the general case of $\nu_e$ transition into any combination 
of $\nu_a$ and $\nu_s$.
It should be noted here that these three measurements do not have
identical energy ranges.  The $SK$ and $SNO$ $CC$ data start from neutrino
energies of 5 and 7 MeV respectively, while the response function
of the $SNO$ $NC$ measurement extends marginally below 5 MeV.  
However the SK rate
and the resulting survival probability show energy independence down
to 5 MeV to a very good precision.  The $SNO$ $CC$ rate shows energy
independence as well, although to lesser precision.  Therefore it is
reasonable to assume a common survival probability for all the three
measurements.  

As recently discussed in, $^6$ the SK elastic and the $SNO$ $CC$ rates can
be combined to access information on $NC$ scattering, so that in
principle there is no new information contained in the $SNO$ $NC$ rate.
In fact eqs. (\ref{six},\ref{seven},\ref{eight}) can be seen to give
the sum rule
\be
R^{NC}_{SNO} = R^{CC}_{SNO} + (R^{el}_{SK} - R^{CC}_{SNO})/r,
\label{nine}
\ee
which predicts $R^{NC}_{SNO} = 1.0 \pm 0.24$ .$^6$  However the large
uncertainty in this prediction reflects the low sensitivity of
$R^{el}_{SK}$ to $NC$ scattering, which is weighted by a small
co-efficient $r$.  On the other hand the expected $NC$ scattering rate
from $SNO$ at the level of $\sim 10^3$ events should have a similar
precision as their present CC rate, i.e. $\pm 8\%$, which will
ultimately go up to $\pm 5\%$.  Keeping in mind the predicted range of
$R^{NC}_{SNO}$ we shall assume 
\be
R^{NC}_{SNO} = 1.0 \pm .08, 1.2 \pm .08,0.8\pm .08
\label{ten}
\ee
as three representative values of the $NC$ scattering rate expected
from $SNO$ and analyse the implications for the solar neutrino
anomaly.  We shall see below that the $R^{NC}_{SNO}$ input will lead
to qualitative improvement in the results of all the three types of
analyses mentioned earlier.
%\bigskip
\begin{figure}[h] %[htbp] %ORIGINAL SIZE: width=1.4TRUEIN; height=1.5TRUEIN
\vspace*{13pt}
\centerline{\psfig{file=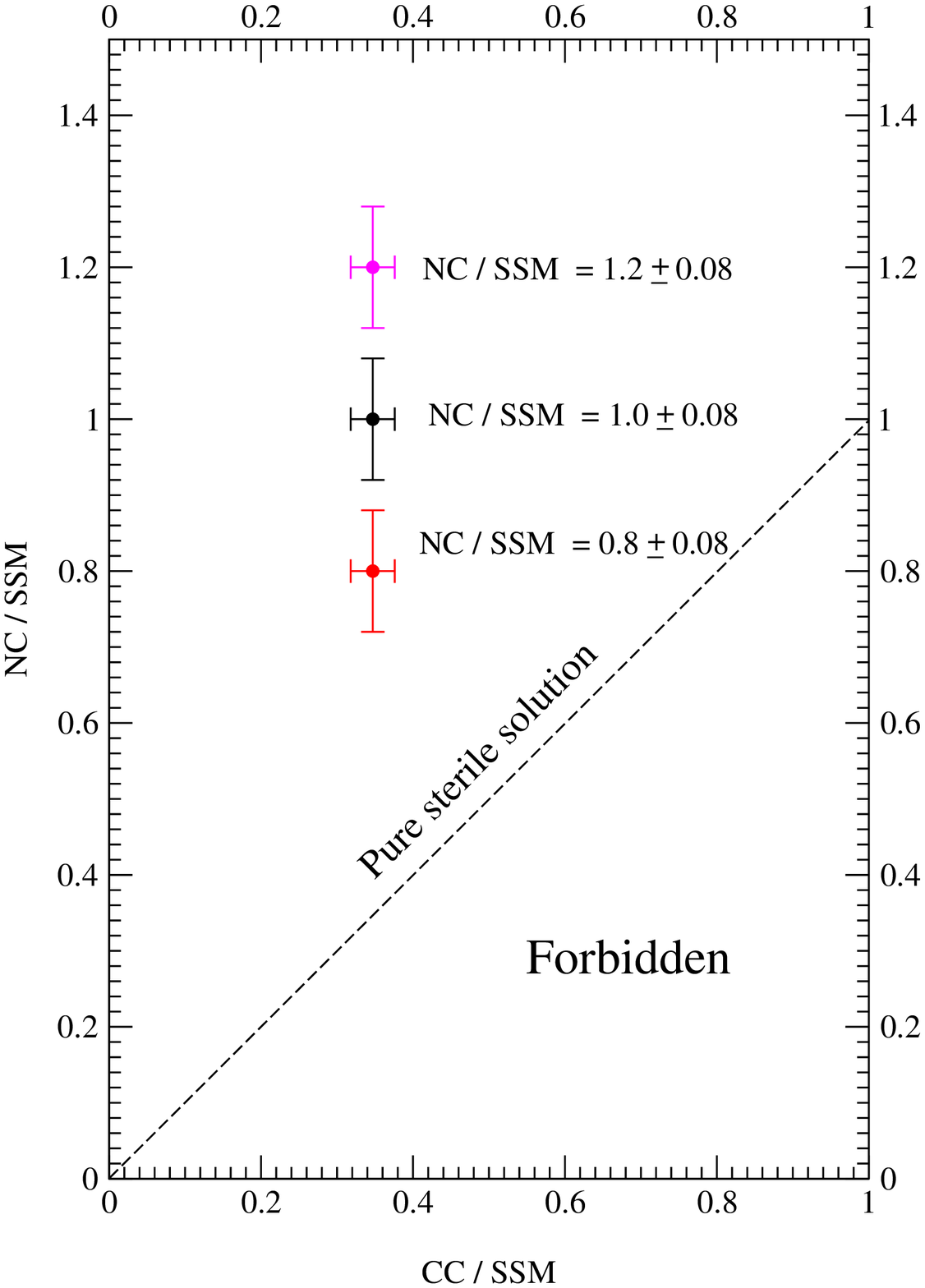,width=10cm,height=12cm}} %100 percent
\vspace*{13pt}
\fcaption{The $SNO$ $CC$ and $NC$ rates shown relative to their $SSM$
predictions for the three representative values of the latter.  The
dashed line is the prediction of the pure $\nu_e$ to $\nu_s$ transition}
\end{figure}

\noindent {\bf General Analysis of $\nu_e \rightarrow \nu_{a,s}$
Transitions}:  Let us start with a model independent analysis of
$\nu_e$ transition into an active neutrino flavour $\nu_a$.  One sees
from eqs. (\ref{six},\ref{seven}) that the observed excess of
$R^{el}_{SK}$ over the $R^{CC}_{SNO}$ in Table 1 constitutes a
$3\sigma$ signal for $\nu_e \rightarrow \nu_a$ 
transition  or
equivalently a $3\sigma$ signal against a pure $\nu_e \rightarrow
\nu_s$ transition .$^4$  Similarly an observed excess of
$R^{NC}_{SNO}$ (\ref{eight}) over $R^{CC}_{SNO}$ (\ref{seven}) will
constitute a model independent signal for the $\nu_e \rightarrow
\nu_a$ transition $P_{ea}$.  Fig. 1 compares the $R^{NC}_{SNO}$ values
of eq. (\ref{ten}) with the current value of $R^{CC}_{SNO}$.  It shows
that an observed value of $R^{NC}_{SNO}$ in the range of $0.8 - 1.2
(\pm .08)$ will constitute of $5-10 \ \sigma$ signal for 
$\nu_e\rightarrow \nu_a$ transition i.e. transition of $\nu_e$ into 
a state of atleast partly active neutrino.

%\begin{figure}
%\begin{center}
%\hspace{10cm}
%\epsfxsize=2in
%\epsfbox{0210fig1.eps}
%\label{fig:0210fig1}
%\caption{The $SNO$ $CC$ and $NC$ rates shown relative to their $SSM$
%predictions for the three representative values of the latter.  The
%dashed line is the prediction of the pure $\nu_e$ to $\nu_s$ transition.}
%\end{center}
%\end{figure}

%\begin{figure}[h] %[htbp] %ORIGINAL SIZE: width=1.4TRUEIN; height=1.5TRUEIN
%\vspace*{13pt}
%\centerline{\psfig{file=0210fig1.eps,width=10cm,height=12cm}} %100 percent
%\vspace*{13pt}
%\fcaption{The $SNO$ $CC$ and $NC$ rates shown relative to their $SSM$
%predictions for the three representative values of the latter.  The
%dashed line is the prediction of the pure $\nu_e$ to $\nu_s$ transition}
%\end{figure}

\begin{figure}[h] %[htbp] %ORIGINAL SIZE: width=1.4TRUEIN; height=1.5TRUEIN
\vspace*{13pt}
\centerline{\psfig{file=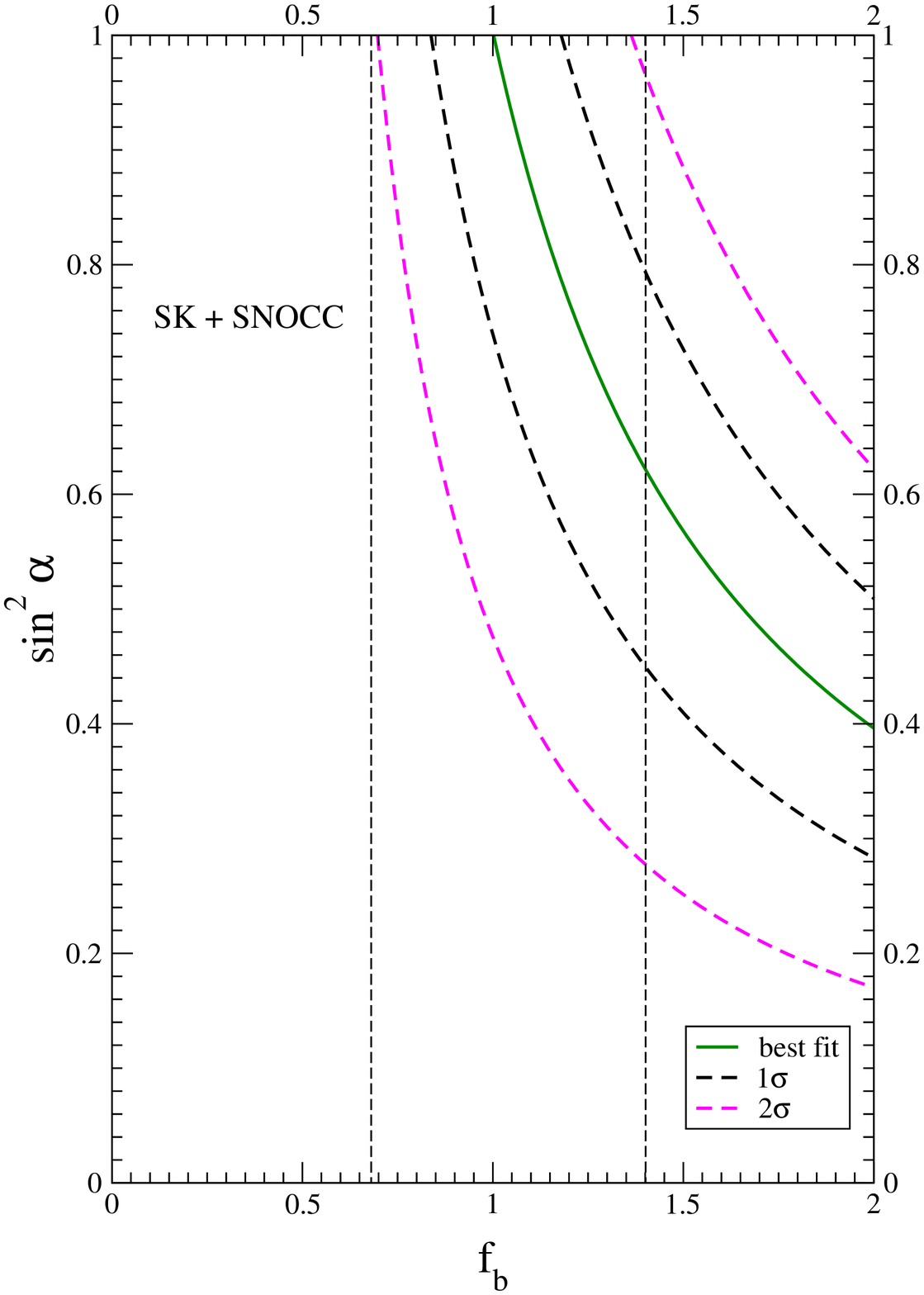,width=10cm,height=10cm}} %100 percent
\vspace*{13pt}
\fcaption{Best fit value of the $^8B$ neutrino flux shown along with
the $1\sigma$ and $2\sigma$ limits against the model parameter
$\sin^2\alpha$,
representing $\nu_e$ transition into a mixed state ($\nu_a \sin\alpha
+ \nu_s \cos\alpha$). The dashed line denote the $\pm 2\sigma$
limits of the $SSM$.}
\end{figure}

\begin{figure}[h] %[htbp] %ORIGINAL SIZE: width=1.4TRUEIN; height=1.5TRUEIN
\vspace*{13pt}
\centerline{\psfig{file=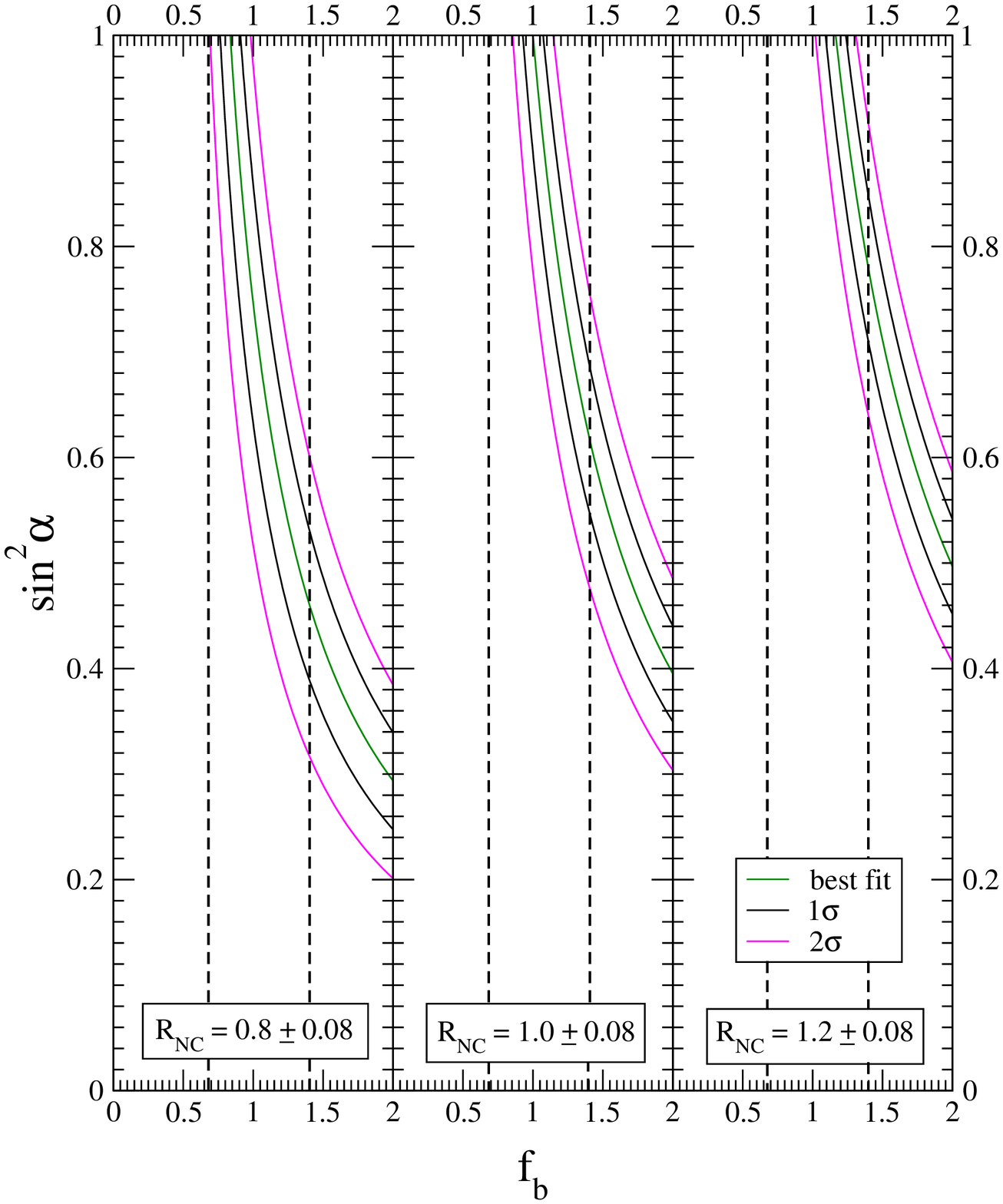,width=14cm,height=10cm}} %100 percent
\vspace*{13pt}
\fcaption{Same as Fig. 2 but including the $NC$ rate to the fit along
with the $SK$ elastic and $SNO$ $CC$ rates.  The dashed lines denote
the $\pm 2\sigma$ limits of the $SSM$.}
\end{figure}

Unfortunately one cannot get a model independent estimate of the
$\nu_e \rightarrow \nu_a$ transition probability $P_{ea}$, since the
above excess corresponds to the product $f_B P_{ea}$.  In fact it is
evident from eqs. (\ref{six}-\ref{eight}) that these two quantities
can not be separated with or without the $P^{NC}_{SNO}$ input.  In
other words for the general case of $\nu_e$ transition into a mixture
of active and sterile neutrinos,
\be
\nu_a \sin\alpha + \nu_s \cos\alpha,
\label{eleven}
\ee
one can rewrite eqs. (\ref{six}) and (\ref{eight}) as 
%\subequations
\label{twelve}
\bea
R^{el}_{SK} &=& f_B P_{ee} + f_B r \sin^2 \alpha (1 - P_{ee}),
\label{twelve-a} \\[2mm]
R^{NC}_{SNO} &=& f_B P_{ee} + f_B \sin^2 \alpha (1 - P_{ee}).
\label{twelve-b}
\eea
%\endsubequations
Then it will not be possible to separately estimate the parameters
$f_B$ and $\sin^2\alpha$.  One can only determine the following
combination in two different ways, i.e.
\bea
\sin^2\alpha (f_B - R^{CC}_{SNO}) &=& (R^{el}_{SK} - R^{CC}_{SNC})/r, 
\label{thirteen} \\[2mm] 
\sin^2\alpha (f_B - R^{CC}_{SNO}) &=& R^{NC}_{SNO} - R^{CC}_{SNO}.
\label{fourteen}
\eea
A two parameter fit to (\ref{thirteen}) was done in ref.$^6$ to
determine the allowed contour in the $f_B - \sin^2\alpha$ plane.  We
shall instead treat $\sin^2\alpha$ as a model
parameter.  And for each input value of $\sin^2\alpha$ we shall
determine $f_B$ first from eq. (\ref{thirteen}) and then from a
weighted average of eqs. (\ref{thirteen}) and (\ref{fourteen}).  The
results are shown as lines corresponding to the central value of $f_B$
along
with the $1\sigma$ and $2\sigma$ boundaries in the $f_B-\sin^2\alpha$ plane in
Figs. 2 and 3.  The advantage of this procedure is that for
$\sin^2\alpha = 1$ the $1\sigma$ and $2\sigma$ ranges of $f_B$ shown
correspond to those of the pure $\nu_e \rightarrow \nu_a$ transition model
without any sterile neutrino, which would not be the case for the two
parameter fit .$^6$

A comparison of Figs. 2 and 3 shows the enormous improvement in
precision one expects by including the $NC$ rate from $SNO$.  The
vertical lines indicate the $2\sigma$ limits on $f_B$ from the SSM.
Combining the two limits one could rule out models having
$\sin^2\alpha < 0.3$ (i.e. singlet neutrino component $\cos^2\alpha >
0.7$) at $2\sigma$ level for $R^{NC}_{SNO} = 0.8 \pm .08$, while
$R^{NC}_{SNO} = 1.2 \pm .08$ would rule out all those singlet neutrino
models with $\sin^2\alpha < 0.6$ ($\cos^2\alpha > 0.4$).  However
within the present uncertainty of the $SSM$ value of the $^8B$
neutrino flux it is unlikely to rule out models with a $\nu_s$
component $\cos^2\alpha < 0.3$.  Nor is it likely to set any upper
limit on $\sin^2\alpha$, which would show incompatibility with pure
$\nu_e \rightarrow \nu_a$ transition.
%\bigskip
\begin{figure}[htbp] %ORIGINAL SIZE: width=1.4TRUEIN; height=1.5TRUEIN
\vspace*{13pt}
\centerline{\psfig{file=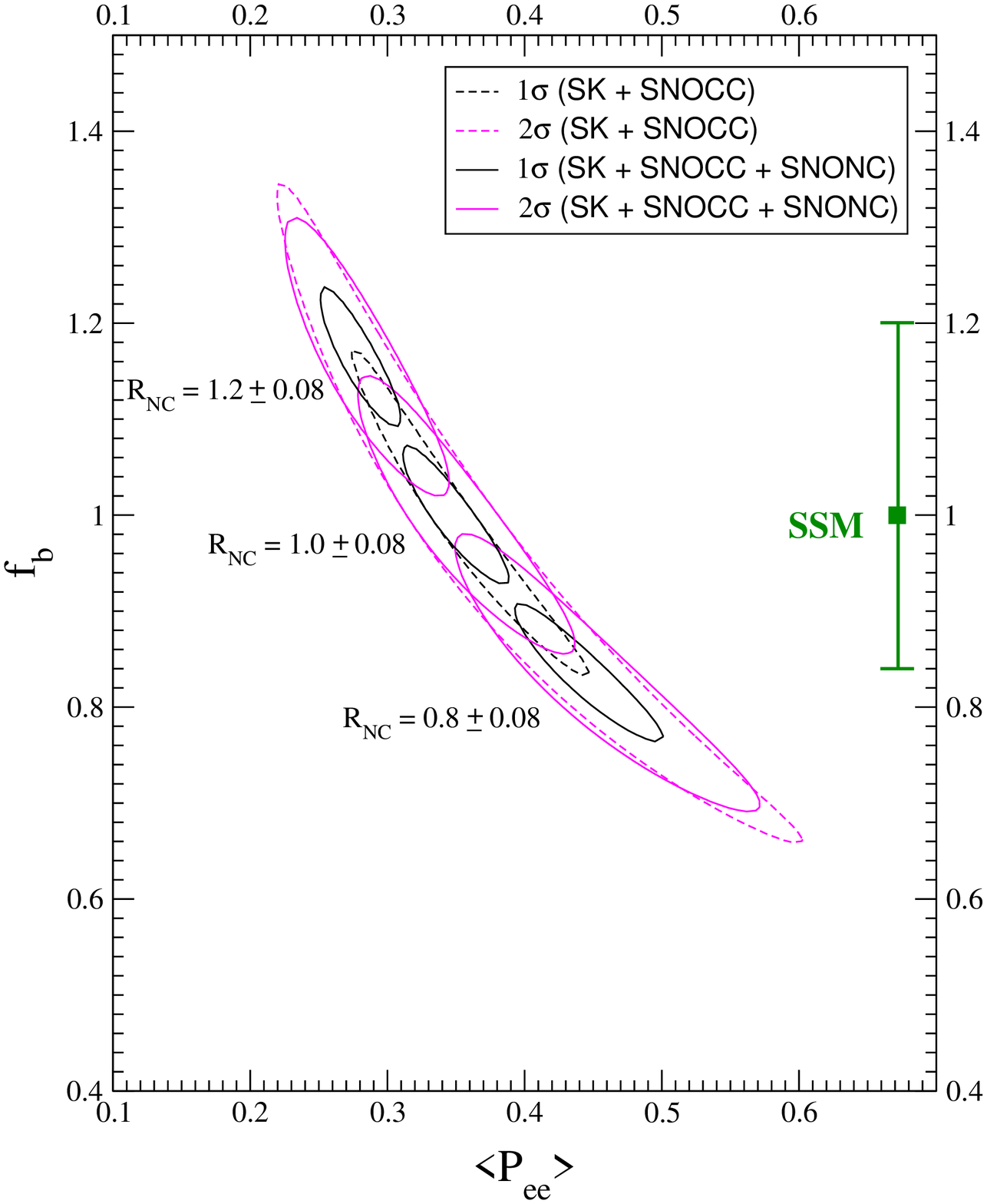,width=14cm,height=12cm}} %100 percent
\vspace*{13pt}
\fcaption{The 1 and $2\sigma$ contours of solutions to the $^8B$
neutrino flux $f_B$ and the $\nu_e$ survival probability $P_{ee}$
assuming $\nu_e$ to $\nu_a$ transition.  The size of the $SSM$ error
bar for $f_B$ is indicated on the right.}
\end{figure}

\noindent {\bf Analysis of pure $\nu_e \rightarrow \nu_a$
Transition}:  Assuming no sterile neutrino the eqs. (\ref{six}) and
(\ref{eight}) reduce to 
%\subequations
\label{fifteen}
\bea
R^{el}_{SK} &=& f_B P_{ee} + f_B r(1 - P_{ee}), 
\label{fifteen-a} \\[2mm] 
R^{NC}_{SNO} &=& f_B.
\label{fifteen-b}
\eea
%\endsubequations
In this case it is possible to estimate both $f_B$ and $P_{ee}$ from
the $SK$ elastic and the $SNO$ $CC$ rates even without the $NC$ rate
from $SNO$ .$^7$  Fig. 4 shows the 1 and 2 sigma contours of this
solution along with the corresponding ones obtained by including the
$R^{NC}_{SNO}$ input.  Our result in the former case is in good
agreement with that of ref. $^7$, although the two methods are not
identical.  Thanks to the effective energy independence of the
survival probability the result is insensitive to the difference
between the energy thresholds of the $SK$ elastic and the $SNO$ $CC$
rates.  Unfortunately the resulting $f_B$ has large error, which
reflects the low sensitivity of $R^{el}_{SK}$ to $f_B$
(eq. \ref{fifteen-a}).  Interestingly both the central value and the
error bar of $f_B$ are practically the same as those of the $SSM$
(shown for comparison near the right scale).  The error in $f_B$
propagates into $P_{ee}$ due to a strong anticorrelation between the
two quantities as their product is well constrained by $R^{CC}_{SNO}$
(\ref{seven}).  Fig. 4 shows that the inclusion of the $NC$ rate from
$SNO$ will result in a reduction in the uncertainty of the
$^8B$ neutrino flux $f_B$ and the corresponding survival probability
$P_{ee}$ by about a factor of 2.5. 
\bigskip

\noindent {\bf The $\nu_e \rightarrow \nu_a$ oscillation
solutions to the Solar Neutrino Anomaly}:  Finally we shall fit the
global solar neutrino data with a standard two-family $(\nu_e
\rightarrow \nu_a)$ oscillation model assuming the $SSM$ fluxes, but
with one difference.  Instead of $R^{el}_{SK}$ and $R^{CC}_{SNO}$ we
shall fit the ratios $R^{el}_{SK}/R^{NC}_{SNO}$ and
$R^{CC}_{SNO}/R^{NC}_{SNO}$. As we see from
eqs. (\ref{seven},\ref{fifteen-a},\ref{fifteen-b}) the $^8B$ neutrino
flux $f_B$ factors out from these two ratios.  Thus the result becomes
immune to the large uncertainty in the $SSM$ value of the $^8B$
neutrino flux (\ref{five}).  We shall include the 19 + 19 day-night
spectral points from $SK$, but with free normalisation to avoid double
counting .$^{7,8,10}$  We shall also include the combined $Ga$ rate of
Table 1 in the fit.  However we shall exclude the $Cl$ rate, since the
experiment has not been independently calibrated.  Besides the
apparent rise of the $CC$ rate between the $Cl$ and $SK/SNO$ energies
is in conflict with the $LMA$ and $LOW$ solutions, which are strongly
favoured by the global fit.$^{10}$
Fig. 5 shows the 90\%, 95\%, 99\% and 99.73\% $(3\sigma)$ $CL$
contours of the fits for the three representative values of
$R^{NC}_{SNO}$ (\ref{ten}), which is the same as $f_B$
(eq. \ref{fifteen-b}).  We have checked that all the three fits have
equally good $\chi^2_{\rm min.} \ (\sim 30)$ and goodness of fit
$(\sim 80\%)$. 

\begin{figure}[htbp] %ORIGINAL SIZE: width=1.4TRUEIN; height=1.5TRUEIN
\vspace*{13pt}
\centerline{\psfig{file=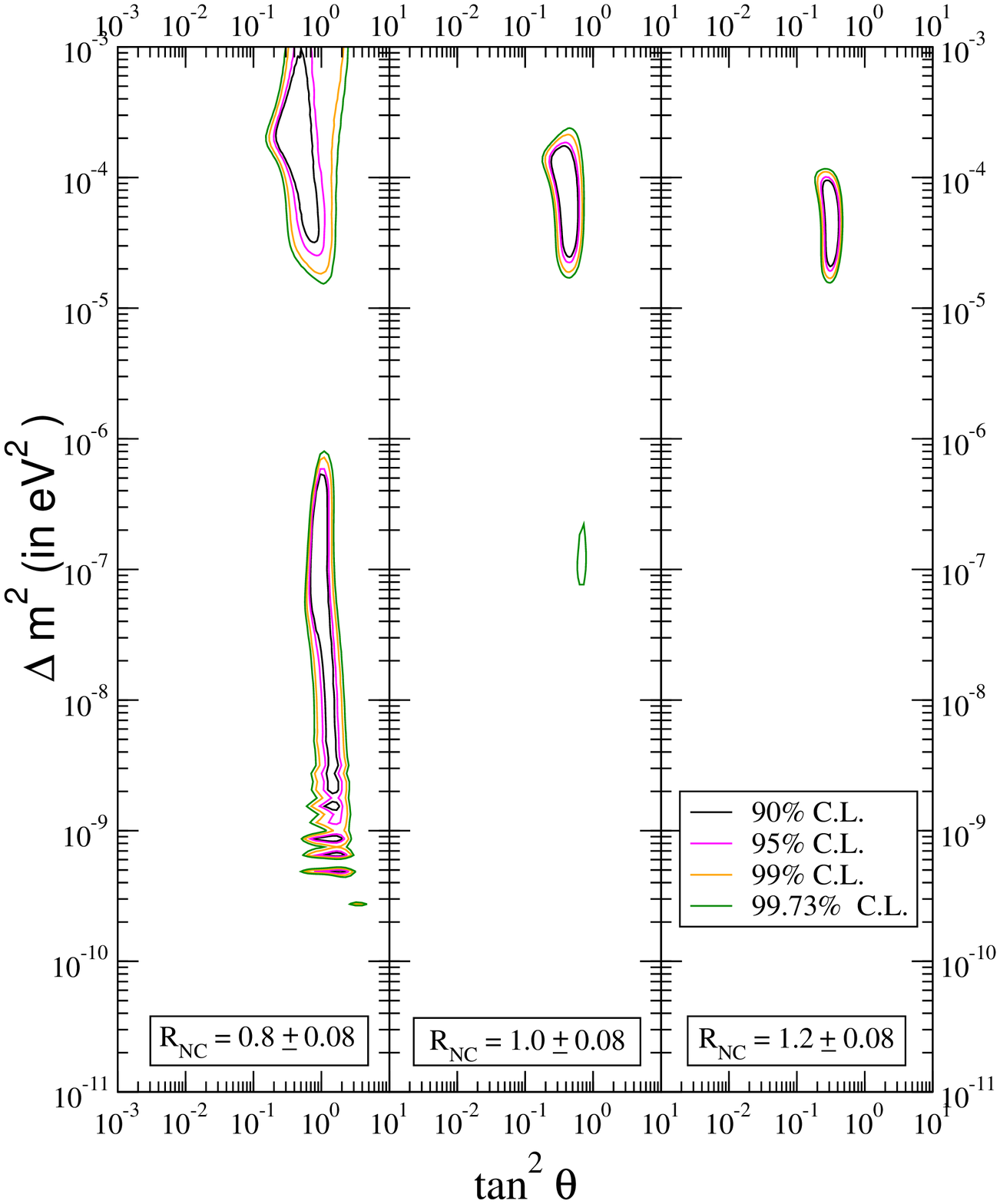,width=14cm,height=16cm}} %100 percent
\vspace*{13pt}
\fcaption{The $\nu_e \rightarrow \nu_a$ oscillation solutions to the
$Ga$ rate, $SK$ day-night energy spectra along with the $SK$ and $SNO$
$(CC)$ rates, both normalised by the $SNO$ ($NC$) rate.}
\end{figure}

The left panel of Fig. 5 represents $R^{SNO}_{NC} = f_B = 0.8 \pm
.08$.  It corresponds to enhancing the survival probability $P_{ee}$
at $SK/SNO$ from 0.35 to nearly 0.45, while $P_{ee}$ at $Ga$ energy
goes up by only 2\%.  Thus the energy dependence of $P_{ee}$ between
the $Ga$ and $SK/SNO$ energies become very mild.  Consequently the
solution favours large mixing angle, going upto maximal mixing, where
it remains valid over two large mass ranges around $\delta m^2 =
10^{-4} \ {\rm eV}^2$ and $10^{-7} \ {\rm eV}^2$ .$^{11}$  These are the
so called $LMA$ and $LOW$ solutions.  In the present context however it
will be more appropriate to call them $HIGH$ and $LOW$ solutions,
since both of them correspond to large mixing angles.  The middle
panel represents $R^{NC}_{SNO} = f_B = 1.0 \pm .08$, which means that
the survival probability $P_{ee}$ at $Ga$ and $SK/SNO$ energies
corresponds to those shown in Table 1.  Since such a large energy
dependence cannot be explained by the earth matter effect, the $LOW$
solution is only allowed at the $3\sigma$ level. On the other hand the
$HIGH$ solution occurs at the upper edge of the $MSW$ region in
$\delta m^2/E$.  Thus the survival probability goes down from
$P_{ee} \simeq 1 - {1\over2} \sin^2 2\theta > 0.5$ above the $MSW$
region to $P_{ee} \simeq \sin^2\theta$ inside it as one moves up from
$Ga$ to $SK/SNO$ energies .$^{10}$  The right panel corresponds to
$R^{NC}_{SNO} = f_B = 1.2 \pm .08$, which means that the survival
probability at $SK/SNO$ energies goes down further by a factor of
1.2.  Consequently the $LOW$ solution disappears completely while the
$HIGH$ solution moves to a lower mixing angle.  Thus the measured
value of the $NC$ rate at $SNO$ can have a profound effect on the mass
and mixing angle of the solar neutrino oscillation solution.
\bigskip

\noindent {\bf Summary}:

\nobreak
In anticipation of the first $NC$ rate from $SNO$ we have analysed how
this data will sharpen our understanding of the solar neutrino
anomaly.  For a quantitative analysis we have chosen three
representative value of this rate, $R^{NC}_{SNO} = 0.8,1.0$ and $1.2
(\pm .08)$.  They span the $\pm 1\sigma$ range of this quantity as
estimated from the $SK$ elastic and $SNO$ $CC$ rates.  The main
results are listed below.
%\begin{alphlist}
\begin{romanlist}
\item It will provide a $5 - 10\sigma$ signal for $\nu_e$
transition into an active flavor $\nu_a$ (or against a pure $\nu_e$
transition into a sterile neutrino $\nu_s$).
\item However for transition into a mixture of $\nu_a$ and
$\nu_s$, we need to know the $^8B$ neutrino flux to constrain the size
of the sterile component.  If we assume the $SSM$ prediction for this
flux then $R^{NC}_{SNO} = 1.2 \pm .08 \ (0.8 \pm .08)$ would imply the
sterile component to be $< 40\%$ (70\%) at the $2\sigma$ level.
\item Assuming a pure $\nu_e \rightarrow \nu_a$ transition one
can combine 
$R^{NC}_{SNO}$ with $R^{CC}_{SNO}$ and $R^{el}_{SK}$ to determine both
the $^8B$ neutrino flux and the $\nu_e$ survival probability to much
higher precision than is possible now from the latter two data.
\item Finally one can do a $\nu_e \rightarrow \nu_a$ oscillation
model fit to the global solar neutrino data assuming the $SSM$ fluxes,
but replacing the $R^{el}_{SK}$ and $R^{CC}_{SNO}$ by the ratios
$R^{el}_{SK}/R^{NC}_{SNO}$ and $R^{CC}_{SNO}/R^{NC}_{SNO}$, which are
independent of the $^8B$ neutrino flux.  For $R^{NC}_{SNO} = 0.8 \pm
.08$ we get both the $LMA$ $(HIGH)$ and $LOW$ solutions covering large
ranges of mass and mixing angle, including maximal mixing.  On the
other hand $R^{NC}_{SNO} \geq 1$ strongly disfavours the $LOW$
solution, while the $HIGH$ solution is restricted to a small patch in
mass and mixing angle excluding maximal mixing.  Thus the measured
value of $R^{NC}_{SNO}$ can have a profound effect on the mass and
mixing parameters of the solar neutrino oscillation.
\end{romanlist}

\nonumsection{Acknowledgments}
\noindent
We thank S. Umasankar for discussions.

\nonumsection{References}

\end{document}